\documentstyle[eqsecnum,epsf,aps,prb,floats]{revtex}

\begin{document}

\twocolumn[\hsize\textwidth\columnwidth\hsize\csname@twocolumnfalse%
\endcsname

\title{The Random-bond Potts model in the large-$q$ limit}

\author{R\'obert Juh\'asz$^{1,2}$, Heiko Rieger$^{3}$, and Ferenc Igl\'oi$^{2,1,4}$}

\address{
$^1$ Institute for Theoretical Physics,
Szeged University, H-6720 Szeged, Hungary\\
$^2$ Research Institute for Solid State Physics and Optics, 
H-1525 Budapest, P.O.Box 49, Hungary\\
$^3$ Theoretische Physik, Universit\"at des Saarlandes, 
     66041 Saarbr\"ucken, Germany\\
$^4$ Centre de Recherches sur les Tr\'es Basses Temp\'eratures, B. P. 166,
F-38042 Grenoble, France
}

\date{April 23, 2001}

\maketitle

\begin{abstract}
  We study the critical behavior of the $q$-state Potts model with
  random ferromagnetic couplings. Working with the cluster
  representation the partition sum of the model in the large-$q$ limit
  is dominated by a single graph, the fractal properties of which are
  related to the critical singularities of the random Potts model.
  The optimization problem of finding the dominant graph, is studied
  on the square lattice by simulated annealing and by a combinatorial
  algorithm.  Critical exponents of the magnetization and the
  correlation length are estimated and conformal predictions are
  compared with numerical results.
\end{abstract}


]

\newcommand{\bc}{\begin{center}}
\newcommand{\ec}{\end{center}}
\newcommand{\be}{\begin{equation}}
\newcommand{\ee}{\end{equation}}
\newcommand{\ba}{\begin{array}}
\newcommand{\ea}{\end{array}}
\newcommand{\beqn}{\begin{eqnarray}}
\newcommand{\eeqn}{\end{eqnarray}}

\section{Introduction}
The effect of quenched disorder at a first-order transition point is
comparatively less understood than the same phenomena at a continuous
transition point. In the latter case relevance-irrelevance criteria,
such as the Harris criterion\cite{harris,chayes} can be used to decide
upon the stability of the pure fixed-point and also perturbation
expansions are developed\cite{perturb} to treat the effect of weak
disorder. If the transition in the pure system is of first order,
neither a general relevance criterion, nor a consistent perturbation
expansion is known to apply around the discontinuity fixed point of
the pure model. One remarkable exception is the stability criterion by
Aizenman and Wehr\cite{aizenmanwehr} (based on an idea of Imry and
Wortis\cite{imrywortis}, see also by Hui and Berker\cite{huiberker}),
which rigorously states that in two dimensions (2d) any amount of
quenched disorder will soften the first-order transition in the pure
system into a continuous one. In 3d the same criterion predicts a
cross-over phenomenon: generally the transition stays discontinuous
for weak disorder, whereas it turns to a second-order one for
sufficiently strong disorder\cite{cbjb}.

Based on the above rigorous results intensive numerical work has
started to clarify the universality class of different disordered
models, which have a discontinuous transition in their pure form. In
2d most of the work has been devoted to the $q$-state Potts model, for
which the transition point is known from self-duality also in its
disordered version\cite{kinzeldomany}, and in the pure model exact
result by Baxter\cite{baxter} ensures a first-order transition for
$q>4$.  Although early Monte Carlo (MC) simulations\cite{earlymc} left
space for an interpretation\cite{kssd} of a $q$-independent
super-universal behavior in random systems, later extensive
MC\cite{pottsmc} and transfer matrix\cite{pottstm} calculations
consistently determined $q$-dependent magnetic exponents, whereas the
correlation length exponent, $\nu$, was found to be close to the pure
Ising value, $\nu_I=1$, for all $q$.

In the large-$q$ limit thermal fluctuations are reduced and as a
consequence the pure model is soluble in any dimension and a
perturbation expansion in powers of $1/q^{1/d}$ can be performed. In
the same limit for the random model at the phase transition point an
effective interface Hamiltonian has been constructed and mapped onto
the interface Hamiltonian of the random-field Ising
model\cite{pottstm}.  This mapping has then been used to relate the
phase diagram of the two problems and to deduce the tricritical
exponents of the RBPM at $d>2$ dimensions. However, in the large-$q$ 
limit no {\it direct} calculation to study the critical
behavior has yet been performed. In 2d the presently known information is
obtained via extrapolation of the results calculated at finite values
for $q$. From these estimates no special type of critical behavior is
expected in the large-$q$ limit. For example the magnetization scaling
dimension, $x_m$, seems to saturate at a finite, non-trivial limiting
value
\cite{olsonyoung,jacobsenpicco} 
$\lim_{q \to \infty} x_m(q) \approx 0.17-0.19$. However, at this point
one should note on the presence of strong (logarithmic) corrections in
the form of $1/\ln q$, see c.f. Fig. 5 in Ref\onlinecite{olsonyoung}.

In the present paper we are going to perform a direct investigation of
the critical behavior of the random-bond Potts model in the large-$q$
limit. As will be shown, in that limit the thermal fluctuations are
negligible and the calculation of the average thermodynamical and
correlation properties of the model is effectively reduced to an
optimization problem. Here the competition between ordering effects, 
originating from a tendency to clustering, and disordering effects, 
due to energy gain from quenched disorder, plays an important role 
in determining the optimal structure. In two dimensions we perform a
numerical study based on simulated annealing and a combinatorial
algorithm, and also conformal aspects of the problem are investigated.

The structure of the paper is the following. In Sect.~2 we introduce
the random cluster representation of the Potts model and define the
equivalent optimization problem emerging in the large-$q$ limit.
Results obtained from the solution of the optimization problem in
different 2d geometries are presented in Sec.~3 and discussed in 
Sec.~4.

\section{Cluster representation in the large-$q$ limit}

We consider the $q$-state Potts model on a $d$-dimensional hyper-cubic
lattice with periodic boundary conditions defined by the Hamiltonian:
\be
\frac{H}{kT}=-\sum_{\langle ij \rangle} K_{ij} \delta(\sigma_i,\sigma_j)\;,
\label{hamilton}
\ee
where $\sigma_i$ are $q$-state Potts variables ($\sigma_i\in\{1,\ldots,q\}$
located at lattice sites $i$, the sum goes over all nearest neighbor pairs 
$\langle ij \rangle$ of the lattice, and $K_{ij}>0$ are reduced
ferromagnetic couplings. The $d$-dimensional hyper-cubic lattice
corresponds to a graph $\overline{G}=(V,E)$, where $V$ is the set of
vertices, which is identical to the lattice sites, and $E$ is the set
of edges, which is identical with the bonds between neighboring sites
on the lattice.  In the random cluster representation\cite{kasteleyn}
the partition sum of the model, $Z$, is expressed as a sum over all
subsets $U\subseteq E$ of the set of edges (or bonds) as:
\be
Z=\sum_{U\subseteq E} q^{n(U)} \prod_{(ij)\in U} v_{ij}\;,
\label{partition}
\ee
where $n(U)$ denotes the number of connected clusters in the subgraph
$G=(V,U)$ of $\overline{G}$, consisting of all lattice sites but the
reduced set of bonds in $U$, and $v_{ij}=e^{K_{ij}}-1$ is the
Mayer-function for the coupling $K_{ij}$. For the latter we use the
parameterization:
\be
v_{ij}=q^{1/d+w_{ij}}\;.
\label{mayer}
\ee
Then the contributions from the different graphs to $Z$ are expressed
in powers of $q$:
\be
Z=\sum_{U\subseteq E} q^{F(U)}
\ee
with
\be
F(U)=n(U)+\sum_{(ij)\in U}(\frac{1}{d}+w_{ij})\;.
\label{costf}
\ee
In the following we consider the large $q$-limit ($q\to\infty$), where
the partition sum is dominated by the leading term given by the
maximum value for $F$:
\be
F_0={\rm max}_{U\subseteq E} \{F(U)\}\;,
\label{cost}
\ee
where $-F_{0}$ corresponds to the free-energy of the system, up to a
pre-factor of $1/(kT\ln q)=const$. Let us denote with $U_0$ the subset
of $E$ that gives the optimum in (\ref{cost}), i.e.~$F_0=F(U_0)$, and
with $G_0=(V,U_0)$ the corresponding {\it dominant graph}. Then the
energetic contribution to $-F_0$ is due to the couplings in the
dominant graph, whereas the entropic term is related to the number of
connected parts. In what follows, we use the word graph when we mean
the subgraph $G=(V,U)$ of $\overline{G}$ defined by an edge subset
$U$.

In the pure system, with $w_{ij}=w$, the structure of the dominant
graphs in the different thermodynamic phases are trivial. Consider
a lattice with $N=L^d$ spins with fully periodic boundary conditions,
the number of bonds is $dN$. Then, in the low-temperature phase with
$w>w_c$ the {\it fully connected graph} $(V,E)$ is the dominant graph,
thus $F_{0}=F_c=[dN(1/d+w)+1]$.  On the other hand in the
high-temperature phase, $w<w_c$, the dominant contribution is due to
the {\it empty graph} $(V,\emptyset)$, with a value of $F_e=N$. At
$w_c=-1/dN$, when $F_c=F_e$, there is phase co-existence, which means
a sharp phase transition even in a finite system in the limit of $q
\to \infty$. In the thermodynamic limit we have $w_c=0$, and the
latent heat per site is given by $\Delta L/N=1$ in our units.

Introducing disorder, such that $w_{ij}$ can take randomly positive
and negative values, the question arises, whether this trivial
structure of the dominant graph persists at the transition point,
i.e.: Is there still a co-existence between two parts of the graph,
one being fully connected, whereas the other is empty? To study this
problem Cardy and Jacobsen\cite{pottstm} have constructed the
interface Hamiltonian, which is then mapped onto that of the RFIM.
This has lead to the answer that for $d>2$ the effect of small
disorder is irrelevant, thus there is still phase co-existence and
thus the transition is of first order, whereas in $d=2$ the phase
co-existence is destroyed by any amount of disorder, in accordance
with Aizenman and Wehr exact results\cite{aizenmanwehr}.

In the following we are going to consider the problem in 2d where the
dominant graph has a non-trivial structure. In particular we study the
(fractal) properties of the largest connected cluster of $G_0$,
denoted by $\Gamma$. In the low-temperature phase, $T<T_c$, $\Gamma$
is compact, thus the average number of points in $\Gamma$ is given by
$[n_{\Gamma}]_{\rm av} \propto N=L^2$, where $L$ is the linear size of
the square lattice and here and in the following $[ \dots
]_{\rm av}$ denotes the average over the quenched disorder. In the
high-temperature phase, for $T>T_c$, $[n_{\Gamma}]_{\rm av}$ stays
finite and defines the average correlation length, $\xi$, through
$[n_{\Gamma}]_{\rm av} \sim \xi^2$. At the transition point the
average mass is expected to scale as
\be
[n_{\Gamma}]_{\rm av} \sim L^{d_f}\;,
\label{frac}
\ee
with a fractal dimension $d_f<2$\cite{note}.

The properties of $[n_{\Gamma}]_{\rm av}$ are directly related to the
asymptotic behavior of the average spin-spin correlation function,
defined in the large-$q$ limit as
\be
[C(r)]_{\rm av}=[\langle \delta(\sigma_i,\sigma_j) \rangle]_{\rm av}\;,
\label{spincorr}
\ee
where $\langle\ldots\rangle$ denotes the thermal and spatial average
over all pairs of sites $i$ and $j$ with a distance $r$. We use
the fact that correlations between two spins are generally zero,
unless they belong to the same cluster, when $C(r)=1$. In the case of
$T \le T_c$, when $\Gamma$ is a spanning cluster the probability,
$Pr(L)$, that a spin belongs to $\Gamma$ is given by
$Pr(L)=[n_{\Gamma}]_{\rm av}/N$, whereas the same probability for two
spins is $Pr(L)^2$. From this follows, that the average correlations
between two spins separated by a large distance of $r=L$ is given by:
$[C(r)]_{\rm av} \simeq Pr(L)^2 =([n_{\Gamma}]_{\rm av}/N)^2$. In the
low-temperature phase, $T<T_c$, where the average magnetization,
$[m]_{\rm av}$, is defined as $[m]_{\rm av}^2=\lim_{r \to
  \infty}[C(r)]_{\rm av}$, we obtain:
\be
[m]_{\rm av}=\lim_{L \to \infty} \frac{[n_{\Gamma}]_{\rm av}}{L^2},
\quad T<T_c\;,
\label{magn}
\ee
whereas at the critical point the average spin-spin correlations decay
as a power:
\be
[C(r)]_{\rm av} \sim r^{-2x_m},\quad x_m=2-d_f,\quad T=T_c\;.
\label{corr_cr}
\ee
Finally, in the high-temperature phase, where the average size of
$\Gamma$ is finite the probability to have a connected cluster of size
$r$ is exponentially small, which leads to an average correlation
function of the form $[C(r)]_{\rm av} \sim \exp(-r/\xi)$, for $r \gg
\xi$.

In the following we specify the form of the disorder, where we make
use of the simplification that arises due to self-duality that holds
under special conditions. According to the results by Kinzel and
Domany\cite{kinzeldomany} the random model is at the critical point,
if the distribution, $P(w)$, of $w_{ij}$ is an even function of $w$,
thus $P(w)=P(-w)$. For convenience we use the bimodal distribution,
\be
P(w)=p \delta(w-\omega)+(1-p)\delta(w+\omega)\;,
\label{bimodal}
\ee
where the critical point is at $p=p_c=1/2$, whereas the reduced
temperature, $t=(T-T_c)/T_c$, can be expressed as:
\be
t=-\omega(p-1/2),\quad |t| \ll 1\;.
\label{reduced_t}
\ee
Generally we restrict ourselves to the range of disorder parameterized
as $0<\omega<1/2$. We note that for $\omega=0$ one recovers the pure
model, whereas for $\omega > 1/2$ we are in the usual percolation
limit. Indeed, for the latter range of parameters the dominant graph
contains all the strong bonds, whereas the weak bonds are all absent.

\section{Methods and results}

According to the results presented in the previous Section the
solution of the RBPM in the large-$q$ limit is equivalent to an
optimization problem with a non-local cost-function given by
Eq.(\ref{costf}).  To find the dominant graph of the problem we used
standard, approximative procedures. Most of the results were obtained
by the method of simulated annealing, but some calculations were
performed by an approximative combinatorial optimization algorithm.

In the procedure of simulated annealing a hypothetical temperature
variable, $T_h$, is introduced and, after thermalization, is lowered
until the hopefully global minimum of the cost-function is reached.
In practical applications we lowered the temperature as
$T_h=1/(\tau-0.5)$, in finite time-steps $\tau=1,2,\dots 60$, and
checked that the resulting configuration does not change after further
cooling.  At a fixed temperature in the thermalization MC steps we
generally used local rules by creating or removing bonds, but
sometimes we also considered to move a full line of bonds. In order to
arrive to the global minimum several different starting configurations
are considered (at least three, sometimes several hundred), and the
best final configuration was taken. In the investigations generally $L
\times L$ finite samples with linear size up to $L=24$ were considered
and periodic boundary conditions were used in both directions. For
smaller sizes the averaging was usually performed over 10000 samples,
whereas for larger sizes we used several thousands of realizations.

Alternatively, for small $\omega$ (precisely for $\omega<0.25$) we
used a combinatorial optimization algorithm that yields a
configuration that is close to the optimum but not necessarily equal
to it. Actually the worst case bound for the ratio of the value
$F_0$ of the optimal solution $U_0$ is to the value $F(U)$
configuration $U$ that is found by the algorithm is only 2/3, which
would be too bad for our purposes.  However, in typical cases the
configurations produced by the algorithm are much closer, as we
checked by comparison with the configurations generated by the
simulated annealing method. The algorithm works as follows
\cite{hagerup}:

For all sites $i$ let $i_{x-}$, $i_{x+}$, $i_{y-}$ and $i_{y+}$ be its
left, right, lower and upper neighbor, respectively, and denote with
$(ii_{x-})$, $(ii_{x+})$, $(ii_{y-})$ and $(ii_{y+})$ the bonds
(edges) between these neighboring sites and $i$. These 
constitute a minimal set of edges that, when removed from
$\overline{G}$ cut the site $i$ from the rest of the graph. Let us
denote them by
\be
E_i:=\{(ii_{x-}),(ii_{x+}),(ii_{y-}),(ii_{y+})\}
\ee
and their weight
\be
w(E_i)=\sum_{(ij)\in E_i} (\frac{1}{2}+w_{ij})\;.
\ee
The minimum cut between any two pairs of sites, $i$ and $j$, (i.e.\ 
the set of edges that has a minimum total weight {\it and} whose
removal from $\overline{G}$ cuts the graph into two disjoint
subgraphs, one containing $i$ and one containing $j$) is then given
either by $E_i$ or $E_j$, as long as $|w_{ij}|<1/4$, as one can easily
convince one-selves.

The idea of the algorithm is as follows: Obviously the removal of the
edges contained in a minimum cut, like in $E_i$ for all $i$, increases
the number of components in the graph by one, i.e. one wins one unit
in the cost function $F(U)$ eq.(\ref{costf}). On the other hand one
looses $w(E_i)$ units and when increasing the number of components of
the graph $G$ one should keep this weight loss as small as
possible. Therefore we consider a collection of minimum cuts as
possible candidates of edge sets to be removed from $\overline{G}$.

Let the edge sets be ordered nondecreasing weight, such that 
$w(E_1)\le w(E_2)\le\ldots\le w(E_{L^2})$ and define for all
$r=0,1,2,\ldots,L^2$ the edge subsets
\be
U_r=E\backslash\bigcup_{i=1}^r E_i\;,
\ee
i.e.\ $U_0=E$, and with increasing $r$ successively edge sets of
non-decreasing weight are substracted from $E$. When doing this
initially (i.e.\ for small $r$) most of the time a site will be
isolated that before has been connected to a larger cluster and
therefore frequently (depending on the weight of the substraced edges)
$F(U)$ will increase, as desired. These are the trial configurations
for our optimization problem and we take the best solution among them,
i.e.\ $U^* $ such that $F(U^*)={\rm max}\{F(U_r)|r=0,1,\ldots,L^2\}$.
It can be shown \cite{hagerup} that $F(U^*)/F(U_{op})\ge2/3$, where $U_{op}$
is the exact optimal solution of (\ref{cost}).
With the combinatorial optimization method we could treat
larger finite systems (up to $128 \times 128$), than by simulated
annealing and the number of configurations we used were between 10000
and 1000 for smaller and larger systems, respectively.

\subsection{Results at the critical point}

First, we tested the relative accuracy of the two methods by comparing
the value of the obtained cost-functions, $F_0$, for different finite
sizes. As a general tendency simulated annealing has given lower,
thus better estimates, but the relative difference for $L \le 16$
was very small, less then  $0.4\%$. For
the largest system we studied by simulated annealing, $L=24$,
the relative difference has increased to about $0.6\%$.
We shall later analyze consequences of the inaccuracy of the min-cut
method in the magnetic properties of the RBPM. In the following illustration
we present results which are obtained by the more accurate simulated
annealing method.

Typical optimal configurations for different values of $\omega$
calculated with the same disorder realization for $w_{ij}(=\pm\omega$)
are presented in Fig.1. The position of the strong bonds
($w_{ij}=+\omega$) can be obtained from the optimal configuration for
$\omega>1/2$, since in percolation only these bonds are occupied. As
seen in the figure for smaller disorder parameter the optimal graph
looks to be more compact, whereas for stronger $\omega$ the optimal
configurations are very close to each other. This fact is a
consequence of the presence of a finite length-scale in the problem.
As shown in the Appendix for small $\omega$ the system behaves
uniformly up to a length-scale, $l_c$, which is estimated as:
\be
l_c \sim \left(\frac{1}{2 \omega}\right)^2\;.
\label{lc}
\ee
To observe the true asymptotic behavior in the RBPM calculation the
system size should be larger than this value, $L>l_c(\omega)$,
therefore we restricted ourselves to not too small $\omega$ values.

\begin{figure}[ht]
\epsfxsize=8truecm
\begin{center}
\mbox{\epsfbox{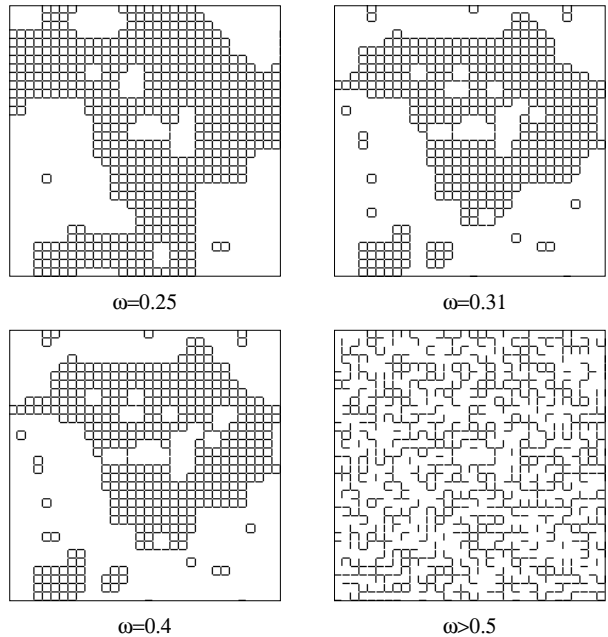}}
\end{center}
\caption{\label{fig1} Typical optimal configurations for different values of
  $\omega$ calculated with the same disorder realization for
  $w_{ij}(=\pm\omega$).}
\end{figure}

Next we analyze the distribution of the largest connected
cluster, $\Gamma$. Inspecting the structure of a typical optimal
graph in Fig.1 we arrive to the conclusion that $\Gamma$ is a fractal,
so that we take the scaling combination $n_{\Gamma}/L^{d_f}$, which
corresponds to the form in Eq.(\ref{frac}). In Fig. 2 we present a
scaling plot of the reduced cluster size distribution, where a data
collapse can be obtained with a fractal dimension of $d_f \approx 1.8$.

\begin{figure}[ht]
\epsfxsize=8truecm
\begin{center}
\mbox{\epsfbox{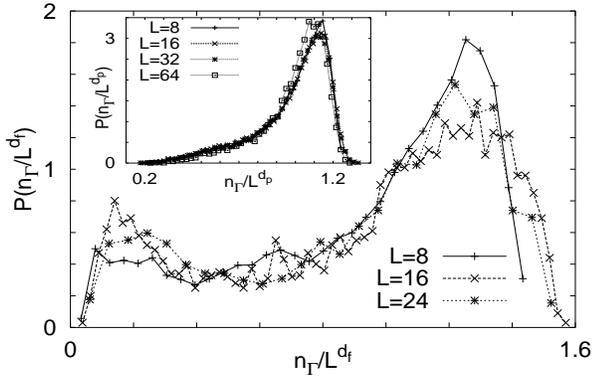}}
\end{center}
\caption{\label{fig2} Scaling plot of the reduced size distribution
of the largest cluster at the critical point of the RBPM at $\omega=0.4$ for different
finite systems. A data collapse is obtained with a fractal dimension of $d_f \approx 1.8$.
In the inset the same quantity is plotted for percolation, when $\omega>1/2$ and
$d_p=91/48$.}
\end{figure}

We note that the points, corresponding to the smallest system, deviate
more from the hypothetical scaling curve, which can be attributed
to the effect of the finite length-scale, $l_c$. In the inset of
Fig.2 a similar scaling plot is presented in the percolation region,
i.e. for $\omega>1/2$, where the fractal dimension of percolation\cite{perc},
$d_p=91/48$ is used. The scaling curves for $\omega<1/2$ and $\omega>1/2$
look different: for the RBPM the distribution is broad and there is
a considerable weight for small clusters, whereas for percolation
the distribution is single peaked without a relevant small
cluster contribution.

Next we calculate the average density of the largest connected
cluster, $[n_{\Gamma}]_{\rm av}/L^2$, from the size dependence of
which the fractal dimension, $d_f$ in Eq.(\ref{frac}) and the
magnetization exponent, $x_m$ in Eq.(\ref{corr_cr}) follows.  In Fig.3
we have plotted $[n_{\Gamma}]_{\rm av}/L^2$ for different finite sizes
in a log-log scale, using different values of the disorder parameter,
$\omega$. In this figure, besides the results obtained by simulated
annealing, also points calculated by the approximate (min-cut)
optimization algorithm are presented. As seen the min-cut algorithm
works satisfactory for small systems, $L \le 16$, when the difference
in the cost-functions calculated by the two methods is also very
small. For larger sizes, however, which are beyond the possibilities
of simulated annealing, the error of the optimization algorithm
increases.  Based on the results presented in Fig.3 the min-cut
method tends to generate a {\it compact cluster} in the large system
limit. Therefore we used the min-cut method only for limited sizes,
which are anyhow manageable by the simulated annealing method,
although with much longer computational time.
\begin{figure}[ht]
\epsfxsize=8truecm
\begin{center}
\mbox{\epsfbox{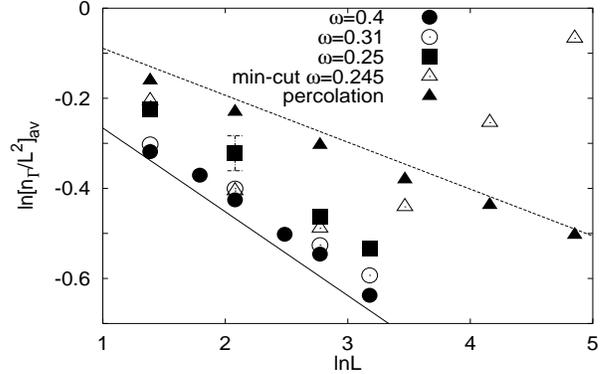}}
\end{center}
\caption{\label{fig3} Size dependence of the average density of the largest
  connected cluster at different values of the disorder parameter,
  $\omega$, calculated by simulated annealing and by the approximative
  optimization (min-cut) algorithm. Note that the min-cut method has a
  systematic error for larger systems. The slope of the curves, $sl$,
  for different $0<\omega<0.5$ is approximately identical and
  indicated by a straight line with $sl=-.2$, but this slope differs
  from that of percolation, which corresponds to $\omega>0.5$, and the
  related straight line has $sl=-5/48$. Typical error of the simulated
  annealing method is indicated by the error bar, whereas the error
  for percolation is smaller than the size of the symbol.}
\end{figure}

Returning to the average density in Fig.3 one can observe that for the
disorder parameter in the RBPM range, i.e. $0<\omega<1/2$, the points
fall on nearly parallel straight lines having a slope of $-2+d_f
\simeq -0.2$, where $d_f \simeq 1.8$ corresponds to the value we used
in the scaling plot of the reduced cluster-size distribution in Fig.
2.  The slope of the same line calculated in the percolation regime,
with $\omega>0.5$ is significantly different, it is $-2+d_p \simeq
-0.1$, where $d_p$ is close to the fractal dimension of 2d percolation.

The estimates of the magnetization scaling dimension, $x_m$, in
Eq.(\ref{corr_cr}) at different disorder parameter, $\omega$, are
summarized in Table I.

\begin{table}
\begin{tabular}{|c||c|}
\hline
$\quad\omega\quad$& $\quad x_m \quad$ \\
\hline
0.2&0.185(30)\\
0.25&0.188(16)\\
0.31&0.165(15)\\
0.4&0.178(13)\\
\hline
$>0.5$&0.103(2)\\
\hline
\end{tabular}
\caption{
Scaling exponent, $x_m$, of the average magnetization for
different disorder parameter, $\omega$. The last row with $\omega>0.5$
corresponds to normal percolation where the exact value is
$x_m^{p}=5/48=0.104$.}
\end{table}

As seen in Table I the average magnetization exponent, $x_m$, is
approximately independent of the disorder parameter for
$0<\omega<1/2$, and its value is within the range of $x_m \approx
0.17-0.19$. This is in agreement with the estimates obtained by
extrapolating the results calculated at finite
$q$-s\cite{olsonyoung,jacobsenpicco}, thus the two limits seem to be
interchangeable. The apparent variation of $x_m$ with $\omega$ can be
attributed to cross-over effects: at $\omega=0$ the pure system
transition, whereas at $\omega=1/2$ the percolation fixed-point is
going to perturb the value of effective, finite-size dependent
exponents.

The average magnetization exponent, $x_m$, has been calculated by
another method, which is based on conformal
invariance\cite{cardy}. Here we use the result, that in a long strip
of width, $L_w$, and with periodic boundary conditions the average
correlation function  along the strip decay exponentially:
\be
[\langle \sigma_i \sigma_{i+u} \rangle]_{\rm av} \sim \exp(-u/\xi_{L_w})\;,
\label{corr_strip}
\ee
where the correlation length, $\xi_{L_w}$, for large widths
asymptotically behaves as:
\be
\xi_{L_w}=\frac{L_w}{2 \pi x_m}\;.
\label{corr_exp}
\ee
In practical calculations we used strips of widths, $L_w=2,3,4$ and $5$,
and with such a lengths that in the calculated correlation function the
exponential decay in Eq.(\ref{corr_strip}) seemed not to change by further increase of
the length. Generally we went at least up to a length of 64 sites, which has
then limited the available widths, $L_w$.
The calculated exponents for some values of the disorder parameter
are given in Table II.

As seen in Table II the size dependence of $x_m$ is very weak for $L_w
\ge 3$ and the extrapolated value of $x_m \simeq 0.17$ is practically
independent of the form of the disorder. This estimate is compatible
with the previous one obtained by finite-size scaling. The fact, that
this latter result lies close to the lower bound of the finite-size
scaling one is probably due to the confluent singularity of the
percolation fixed-point, which is quite strong in the region of
$\omega$-s we used in the calculation on strips.

\begin{table}
\begin{tabular}{|c||c|c|c|c|}
\hline
$\quad\omega\quad$ & \multicolumn{4}{c|}{$x_m$}\\
\hline
& $\quad L_w=2\quad$ & $\quad L_w=3\quad$ & $\quad L_w=4\quad$& $\quad L_w=5\quad$\\
\hline
0.400&0.263(9)&0.166(4)&0.165(5)&0.163(6)\\
0.423&0.267(1)&0.168(5)&0.167(2)&0.163(6)\\
0.452&0.266(1)&0.170(4)&0.169(2)&0.163(6)\\
\hline
\end{tabular}
\caption{
Numerical estimates for the average magnetization exponent,
$x_m$, using the correlation length-exponent relation in Eq.(\ref{corr_exp})
for different widths, $L_w$.}
\end{table}

We have also calculated the central charge of the conformal anomaly, $c$,
from the finite-size correction to the free-energy per width:
\be
f_0(L_w)=f_0(\infty)-\frac{\pi c}{6 L_w^2}+ {\cal O}(L_w^{-3})\;,
\label{c_anomaly}
\ee
with the result:
\be
c=0.74(1)=\frac{0.51(1)}{\ln 2}\;.
\ee
This is compatible with previous estimate\cite{jacobsenpicco}
$c\simeq 0.5/\ln2$, which is obtained by finite-$q$ extrapolation.

\subsection{Results outside the critical point}

We close our paper by an investigation of the average magnetization,
$[m(L,t)]_{\rm av}$, in the vicinity of the critical point. In the
scaling region, defined as $L|t|^{\nu}=O(1)$, where $\nu$ is the
critical exponent of average correlations, the average magnetization
is expected to behave as:
\be
[m(L,t)]_{\rm av}=L^{-x_m} \tilde{m}(L|t|^{\nu})\;,
\label{m_scaling}
\ee
where $\tilde{m}(y)$ is some scaling function.  The calculated
magnetizations at different finite size and temperature then should
collapse to the same scaling function, provided the correct critical
exponents, $x_m$ and $\nu$ are used. 
\begin{figure}[ht]
\epsfxsize=8truecm
\begin{center}
\mbox{\epsfbox{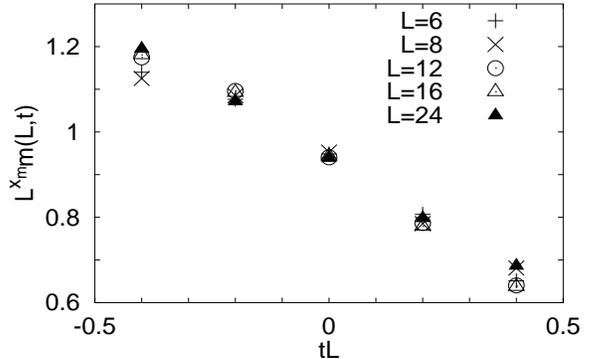}}
\end{center}
\caption{\label{fig4} Scaling plot of the finite-size average magnetization
in the vicinity of the critical point, for a disorder parameter $\omega=0.4$.
The scaling exponents we used here are $x_m=0.177$ and $\nu=1$.}
\end{figure}
In Fig.4 we show the result of
such a scaling plot, where we used $\nu=1$, as found approximately in
finite-$q$ calculations, whereas for $x_m$ we used our previous
estimate obtained through finite-size scaling at the critical
point. The data collapse in Fig.4 is satisfactory, however to obtain a
precise estimate on $\nu$ one needs to extend the calculations for
larger systems.

\section{Discussion}
In this paper the critical behavior of the Potts model with non-frustrated,
random couplings is studied in the large $q$-limit. We have shown how the
calculation of the free energy and the correlation functions of the RBPM
can be mapped onto an optimization problem, which is then studied by
simulated annealing and by an approximate combinatorial optimization
algorithm. Working with the bimodal distribution in Eq.(\ref{bimodal})
our results are compatible with the RG-phase diagram drawn in Fig.5.
\begin{figure}[ht]
\epsfxsize=8truecm
\begin{center}
\mbox{\epsfbox{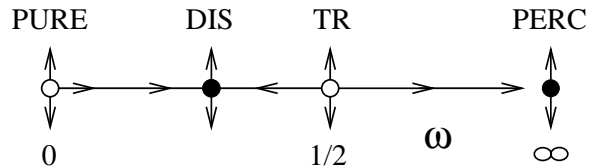}}
\end{center}
\caption{\label{fig5} Schematic RG phase diagram of the 2d RBPM with varying
strength of bimodal disorder, $\omega$. For details see the text.}
\end{figure}

The pure systems fixed-point ($PURE$), situated at $\omega=0$, is unstable
against any weak disorder, thus the critical behavior in the range
of $0<\omega<0.5$ is controlled by the disordered fixed point
($DIS$). Our numerical calculation indeed indicate a universality with
respect of the strength of disorder.  Increasing the disorder over
$\omega=0.5$ we reach the region of attraction of the normal
percolation, and the corresponding fixed point ($PERC$) is located at
$\omega=\infty$. Our RG phase diagram is completed by introducing a repulsive
tricritical fixed point, $TR$, at $\omega=0.5$, which separates
the regions of attraction of the two non-trivial fixed points,
$DIS$ and $PERC$. The singular properties of the $TR$ can be
quite unusual, since the corresponding optimal graph is highly degenerate:
the possible configurations include all which
interpolate between that of the RBPM and that
of normal percolation.

The behavior of the system at the fixed-point $DIS$, which is the subject
of the present paper, is strongly dominated by disorder effects, whereas
thermal fluctuations seem to be negligible. Similar, disorder dominated
critical behavior occur in random quantum spin chains, where analytical
results are available\cite{fisher,bigpaper,ijl01},
and also in 2d random quantum ferromagnets\cite{2dRG}. Whether
exact results can be obtained also for the 2d RBPM in the large-$q$
limit will be seen in future research.

Acknowledgment:
F.I. is grateful to J-C. Angl\'es d'Auriac and  L. Turban
for useful discussions. This work has been
supported by a German-Hungarian exchange program (DAAD-M\"OB), by the Hungarian
National Research Fund under grant No OTKA
TO23642, TO25139, TO34183 and  MO28418 and by the Ministry of Education under
grant No FKFP 87/2001.

\appendix
\section*{Length-scale in the small disorder limit}

Here we estimate the size, $l$, of a step, which is situated at the top
of a straight surface of a connected cluster, see Fig. 6.
\begin{figure}[ht]
\epsfxsize=8truecm
\begin{center}
\mbox{\epsfbox{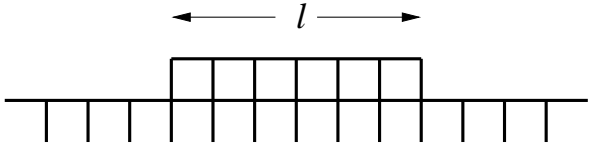}}
\end{center}
\caption{\label{fig6} A connected cluster with a step of $l$-points on the top of
a straight surface.}
\end{figure}
Using the bimodal distribution in Eq.(\ref{bimodal}) the existence of the step is
connected to the condition:
\be
\sum_{i=1}^{2l-1} \left(\frac{1}{2}+w_i \right) > l\;,
\ee
where $w_i=\pm \omega$ with the same probability, or equivalently:
\be
\sum_{i=1}^{2l-1} p_i > \frac{1}{2 \omega}\;,
\label{sum}
\ee
with $p_i= \pm 1$. For large $l$ the probability distribution of the sum
in the l.h.s. of Eq.(\ref{sum}) is Gaussian, with a variance of $\sqrt{2l-1}$.
Consequently the average size of the step, $l_c$, scales with a small $\omega$
as
\be
l_c \sim \left(\frac{1}{2 \omega} \right)^2\;,
\ee
as given in Eq.(\ref{lc}).


\begin{references}

\bibitem{harris}
        A.B. Harris, J. Phys. C{\bf 7}, 1671 (1974).

\bibitem{chayes}
        J. T. Chayes, L. Chayes, D. S. Fisher and T. Spencer,
        Phys. Rev. Lett. {\bf 57}, 299 (1986).

\bibitem{perturb}
        A.W.W. Ludwig, Nucl. Phys. B {\bf 285} [FS19], 97 (1987);
        Nucl. Phys. B{\bf 330} [FS19], 639 (1990); A.W.W. Ludwig and
        J.L. Cardy, Nucl. Phys. B{\bf 330} [FS19], 687 (1990);
        Vl. Dotsenko, M. Picco and P. Pujol, Nucl. Phys. B{\bf 455} [FS], 701 (1995);
        M.A. Lewis, Europhys. Lett. {\bf 43}, 189 (1998).

\bibitem{aizenmanwehr}
        M. Aizenman and J. Wehr, Phys. Rev. Lett. {\bf 62}, 2503 (1989).

\bibitem{imrywortis}
        Y. Imry and M. Wortis, Phys. Rev. B{\bf 19}, 3580 (1979).

\bibitem{huiberker}
        K. Hui and A.N. Berker, Phys, Rev. Lett. {\bf 62}, 2507 (1989).

\bibitem{cbjb}
        For a recent numerical study about the 3d random Potts model see: C. Chatelain,
        B. Berche, W. Janke and P.E. Berche, cond-mat/0103377. 

\bibitem{kinzeldomany}
        W. Kinzel and E. Domany, Phys. Rev. B{\bf 23}, 3421 (1981). 

\bibitem{baxter}
        R.J. Baxter, J. Phys. C{\bf 6}, L445 (1973).

\bibitem{earlymc}
        S. Chen, A.M. Ferrenberg, and D.P. Landau, Phys. Rev. Lett.{\bf 69}, 1213 (1992);
        Phys. Rev. E{\bf 52}, 1377 (1995); E. Domany and S. Wiseman, Phys. Rev. E{\bf 51},
        3074 (1995).

\bibitem{kssd}
        M. Kardar, A.L. Stella, G. Sartoni and B. Derrida, Phys. Rev. E{\bf 52}, R1269 (1995).

\bibitem{pottsmc}
        M. Picco, Phys. Rev. Lett. {\bf 79}, 2998 (1997); C. Chatelain and B. Berche,
        Phys. Rev. Lett. {\bf 80}, 1670 (1998); Phys. Rev. E{\bf 58} R6899 (1998);
        {\bf 60}, 3853 (1999); G. Pal\'agyi, C. Chatelain, B. Berche and F. Igl\'oi,
        Eur. Phys. J. B{\bf 13}, 357 (2000). 

\bibitem{pottstm}
        J.L. Cardy and J.L. Jacobsen, Phys. Rev. Lett. {\bf 79}, 4063 (1997),
        J.L. Jacobsen and J.L. Cardy, Nucl. Phys. B{\bf 515}, 701 (1998).       

\bibitem{olsonyoung}
        T. Olson and A.P. Young, Phys. Rev. B{\bf 60}, 3428 (1999).

\bibitem{jacobsenpicco}
        J.L. Jacobsen and M. Picco, Phys. Rev. E{\bf 61}, R13 (2000).

\bibitem{kasteleyn}
        P.W. Kasteleyn and C.M. Fortuin, J. Phys. Soc. Jpn. {\bf 46} (suppl.),
        11 (1969).

\bibitem{note}
        At the critical point the dominant graph is generally not unique, c.f. with
        the bimodal distribution in Eq.(\ref{bimodal}) any two clusters having just one
        strong and one weak bond between them could be either connected or disconnected.
        We assume that the degenerate optimal graphs have the same asymptotic fractal
        properties.

\bibitem{hagerup}
         T. Hagerup, unpublished.

\bibitem{cardy}
        See e.q. J.L. Cardy, in {\it Phase Transitions and Critical Phenomena, Vol. 11, p. 1},
        edited by J.L. Lebowitz and M.S. Green, (London, Academic) (1987).

\bibitem{perc}
        D. Stauffer and A. Aharony, {\it Introduction to Percolation Theory},
        (London: Taylor $\&$ Francis) (1992).

\bibitem{fisher}
        D.S. Fisher, Phys. Rev. Lett. {\bf 69}, 534 (1992); 
        Phys. Rev. B {\bf 51}, 6411 (1995).

\bibitem{bigpaper}
        F. Igl\'oi and H. Rieger, Phys. Rev. B{\bf 57} 11404 (1998).

\bibitem{ijl01}
        F. Igl\'oi, R. Juh\'asz and P. Lajk\'o, Phys. Rev. Lett. {\bf 86}, 1343 (2001).

\bibitem{2dRG}
        O. Motrunich, S.-C. Mau, D.A. Huse and D.S. Fisher, Phys. Rev. B{\bf 61}, 1160 (2000);
        Y-C. Lin, N. Kawashima, F. Igl\'oi and H. Rieger, Prog. Th. Phys. (Suppl.) {\bf 138},
        470 (2000).

\end{references}
\end{document}